\documentclass[twocolumn,aps,english,prb,showpacs,longbibliography]{revtex4-1}

\usepackage{natbib}
\usepackage{graphicx}

\usepackage{color}
\usepackage{commath}
\usepackage{graphicx}
\usepackage{bm}
\usepackage{verbatim}

\begin{document}

\title{Surface anisotropy in a magnetic cylinder induced by the displacement of a vortex core}
\author{A. Riveros$^1$}
\author{D. A. Carvajal$^2$}
\author{J. Escrig$^{1,3}$}

\affiliation{$^{1}$Departamento de F\'{\i}sica, Universidad de Santiago de Chile (USACH),
Av. Ecuador 3493, 9170124 Santiago, Chile\\
$^{2}$Departamento de Matem\'{a}tica y Ciencia de la Computaci\'{o}n, Universidad de Santiago de Chile (USACH), Las Sophoras 173, 9170020 Santiago, Chile\\
$^{1,3}$Center for the Development of Nanoscience and Nanotechnology (CEDENNA), 9170124 Santiago, Chile
}

\begin{abstract}
In this article we investigate the induction of a surface anisotropy due to the displacement of the vortex core in a cylindrical nanostructure. In fact, the effect of the displacement of the vortex core in the dipolar energy can be modeled simply as a surface anisotropy of the form $E_s = K_s \int_{\mathcal{S}_m} d\mathcal{S} \, (\hat{n} \cdot \hat{m})^2/2$. Moreover, the surface anisotropy constant $K_s$ is proportional to the cylinder in-plane demagnetizing factor in the direction of the core deviation, $N_y(L/R)$, i.e., $K_s = \mu_0 M_0^2 R \,  N_y(L/R)$, where $R$ and $L$ are the radius and the thickness of the cylinder, respectively. Our results show that the term of the nontrivial dipolar energy caused by the charges in the cylinder mantle can be replaced by a simple integral $E_s$ that increases the efficiency of the numerical calculations in the analytical study of the displacement of the vortex core in magnetic vortices.

\end{abstract}

\maketitle

\section{Introduction}

Magnetic vortices are magnetic textures composed of a curling in-plane magnetization that turns out of plane at its center over a  region known as the vortex core \cite{SOH+00,WWB+02}. These magnetic textures are very attractive both from a fundamental and applied point of view \cite{MAL+10,IFY+12,VPS+06,VCW+09,MCK+14,BSB+14}, because they have no significant magnetic stray-fields, crosstalk between magnetic elements is avoided, enabling close packing required for potential technological applications \cite{YHY+13,PdK+10,PdK+11,YJL+11,UUH+13}. The magnetic vortices can be stabilized in cylindrical nanostructures, due to the effects of edges, converting these nanostructures into potential candidates for nonvolatile memories \cite{YKN+07,KWC+10}, biomedicine \cite{KRU+10} and new logic operators \cite{JCL+12}. 

On the other hand, the gyrotropic mode \cite{CAS+04,GIN+02,PEE+03,UK02,IZ02}, which is the lowest frequency excitation of a magnetic vortex, is very different from the precessional modes that are typically seen in uniformly magnetized cylinders \cite{GIN+02,Thiele73}. In this case, the vortex core makes an orbit-like trajectory around its equilibrium position, enabling a description of the mode using a collective co-ordinate approach. The girotropic mode of a cylindrical nanostructure can be excited by the application of pulsed or oscillating in-plane \cite{GIN+02,UK02,GHK+06,KLC+08} and out-of-plane \cite{FFK+16} magnetic fields, as well as by the application of spin-polarized electric currents (mechanism known as spin transfer torque \cite{SKT+06,PKF+07,DGG+10,KBV+10,HK11}), with the mode's resonant frequency depending on the vortex magnetization configuration \cite{GIN+02}. The possibility of controlling the movement of the vortex core by fields or currents can be exploited for applications in electronic oscillators \cite{PKF+07,DGG+10,LLT+14}, frequency-controlled data storage \cite{PdK+10}, and frequency-based \cite{BGW+10} magnetic field detection \cite{FM16,WBW+15}. 

It is well known that in the case of cylindrical nanostructures there are two effective anisotropies: one is related to the surface charges that appear on the faces of the cylinder, which favors an easy-plane anisotropy for thin cylinders and an easy-axis anisotropy for thick cylinders, and another related to the surface charges that appear on the edge of the cylinder, which is responsible for the tangential magnetization distribution along the cylinder edge resulting in clockwise or counterclockwise vortex chirality \cite{PSK+14}. In general, the surface anisotropies of the different nanostructures induce specific magnetic configurations \cite{LDB+07}. However, in this article we are interested in investigating the opposite effect, that is, the induction of a surface anisotropy due to the displacement of the vortex core in a cylindrical nanostructure.

\section{Theoretical model}

From Fig. 1 it can be seen that if a magnetic field is applied perpendicular to the axis of the cylinder (for example, in the $x$ direction), the vortex core moves a distance $s$ in the direction perpendicular to the field (in the example, to the $y$ direction) \cite{SHZ00,GM01}. While this displacement is favored by the exchange and Zeeman energies, it is hindered by the dipolar energy, $E_d$ \cite{GNO+01}. By considering a rigid vortex and choosing the reference system in the center of the displaced core maintains the azimuthal symmetry, and the behavior of the magnetization is a function of the radial distance with respect to the center of the core, so that only the surface charges contribute to the dipolar energy, i.e., $E_d(s) = E_c + E_m(s)$, where $E_c$ is associated to the magnetostatic potential, $U_c$, related with the surface charges of the cylinder covers, which does not depend on the deviation $s$, while $E_m(s) = \mu_0 M_0 \int_V dV\, \hat{m}(\vec{r})\cdot \vec{\nabla} U_m(\vec{r})/2$ corresponds to the dipolar energy associated with the surface charges of the mantle of the cylinder, with a magnetostatic potential given by $U_m = \int_{\mathcal{S}_m}  d\mathcal{S}'\, \hat{n}'\cdot \vec{M}(\vec{r}\,')/(4 \pi \, |\vec{r}-\vec{r}\,'|)$, where $\mathcal{S}_m$ is the surface of the mantle of the cylinder. Thus,
\begin{figure}[ht]
\centering
\includegraphics[width=0.6\linewidth]{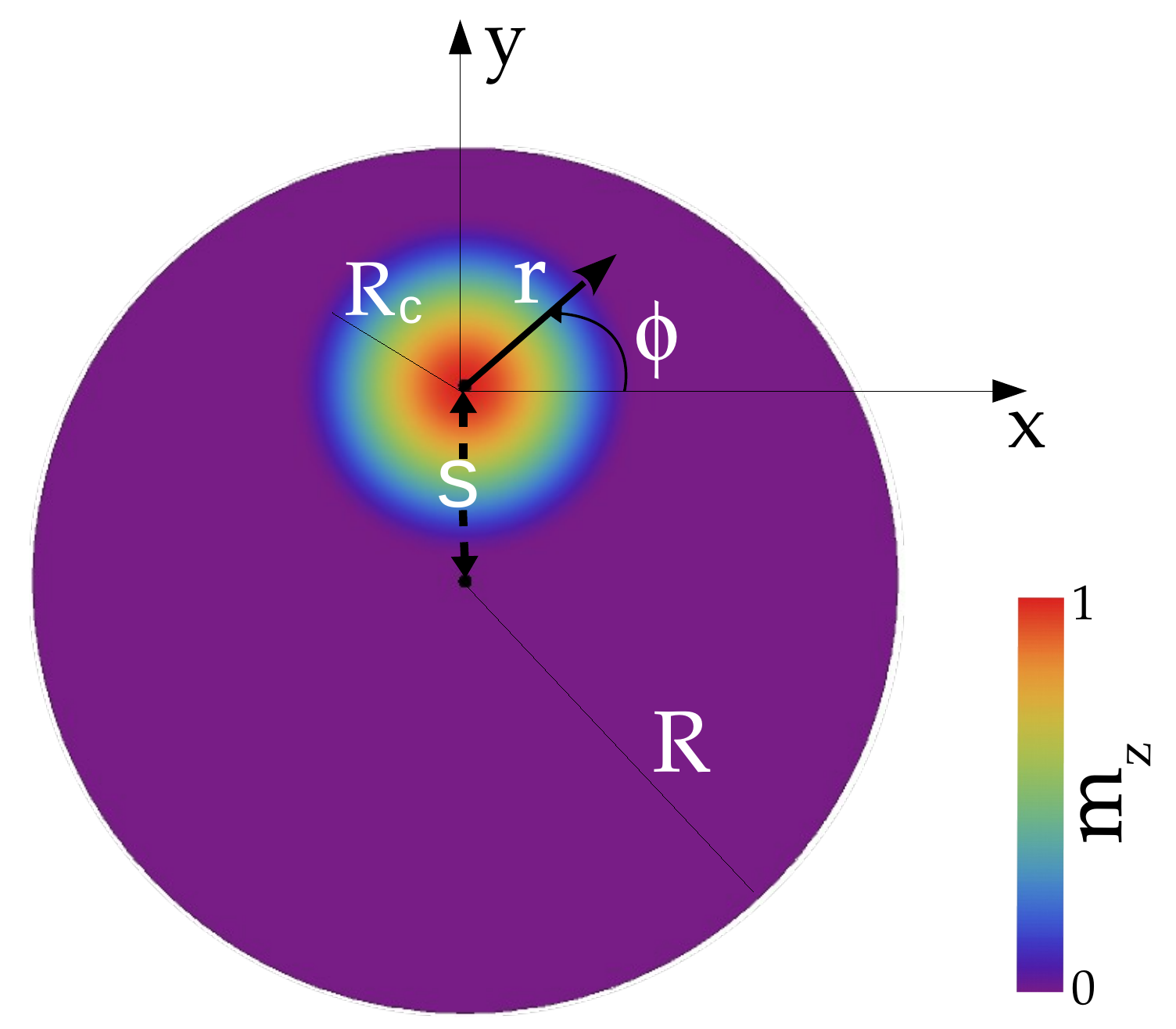}
\caption{(Color online) Top view of a vortex with its core displaced a distance $s$ from its equilibrium position. The cylinder has a radius $R$ and a thickness $L$. The magnetization $m_z(r)$ is presented as a color density plot for the ansatz obtained from Ref. [3].}
\label{picture}
\end{figure}
\begin{eqnarray}
\label{Um}
U_\text{m} = \frac{M_0 \mathcal{C}}{4 \pi} \frac{s}{R} & \displaystyle{\int_0^{2 \pi}} \hspace{-0.1cm} d\phi' \mathcal{J}(\phi',s) \cos{\phi'} \left[ \ln{(z + \sqrt{\mathcal{A}+z^2})} \right. \nonumber \\
 &\left. -\ln{(z-L + \sqrt{\mathcal{A} +(L-z)^2})} \right]
\end{eqnarray}
and
\begin{eqnarray}
\label{Edm}
&E_m (s) =  \displaystyle{\frac{\mu_0 M_0^2}{4\pi}\frac{s}{R} \int_0^{2\pi}} \hspace{-0.1cm}d\phi \int_{0}^{r_\text{max}(\phi,s)} \hspace{-0.5cm} dr r\int_0^{2\pi} \hspace{-0.1cm} d\phi'\mathcal{J}(\phi',s)   \nonumber\\
&\times r_\text{max}(\phi',s) \cos{\phi'} \sin{(\phi-\phi')} \displaystyle{\left[ \frac{\sqrt{\mathcal{A}}-\sqrt{\mathcal{A}+L^2}}{\mathcal{A}} \right]} \, , 
\end{eqnarray}
where we have considered the deviation of the core in the $y$-direction (as shown in Fig. \ref{picture}), and have defined $\mathcal{J}(\phi,s) =\sqrt{r_\text{max}^2(\phi,s)+ (\partial_{\phi} r_\text{max}(\phi,s))^2} $, $r_\text{max}(\phi,s) = -s \sin{\phi} + \sqrt{R^2 - s^2 \cos^2{\phi}}$ and $\mathcal{A} = (r \, \hat{r}- r_\text{max}(\phi',s)\, \hat{r}')^2 $. The integral expression given in Eq. \eqref{Edm} is equivalent to the infinite series expansion of Ref. [40], $E_m^{(\mu)}(s)= \sum_{i=1}^{\mu} \mathcal{E}_m^{(i)}(s)$ in the limit $\mu \rightarrow \infty$. The convergence of $E_m^{(\mu)}(s)$ as $\mu$ increases to the limit $E_m^{(\infty)}(s) = E_m(s)$ is shown in Fig. \ref{covergence_Edm} by plotting $E_m^{(\mu)}(s)$ as a function of the core deviation aspect ratio $s/R$, for $\mu = 1,2,10,50$ and $\mu \rightarrow \infty$, for a dot of $R = 100$ nm in radius and a) $L = 10$ nm and b) $L =50$ nm in thickness. The curve for the limit $\mu \rightarrow \infty$ was obtained using Eq. \eqref{Edm}. The insets show a zoom of the curves for $0.98 \leq s/R \leq 0.985$. Since the core deviation length is restricted by $s < R - R_c$, the limit $s = R$ (where the magnetic energy is singular \cite{GM01,GNO+01} which should be treated more carefully \cite{MKN+13}) is not reached in this work.
\begin{figure}[ht]
\centering
\includegraphics[width=1.0\linewidth]{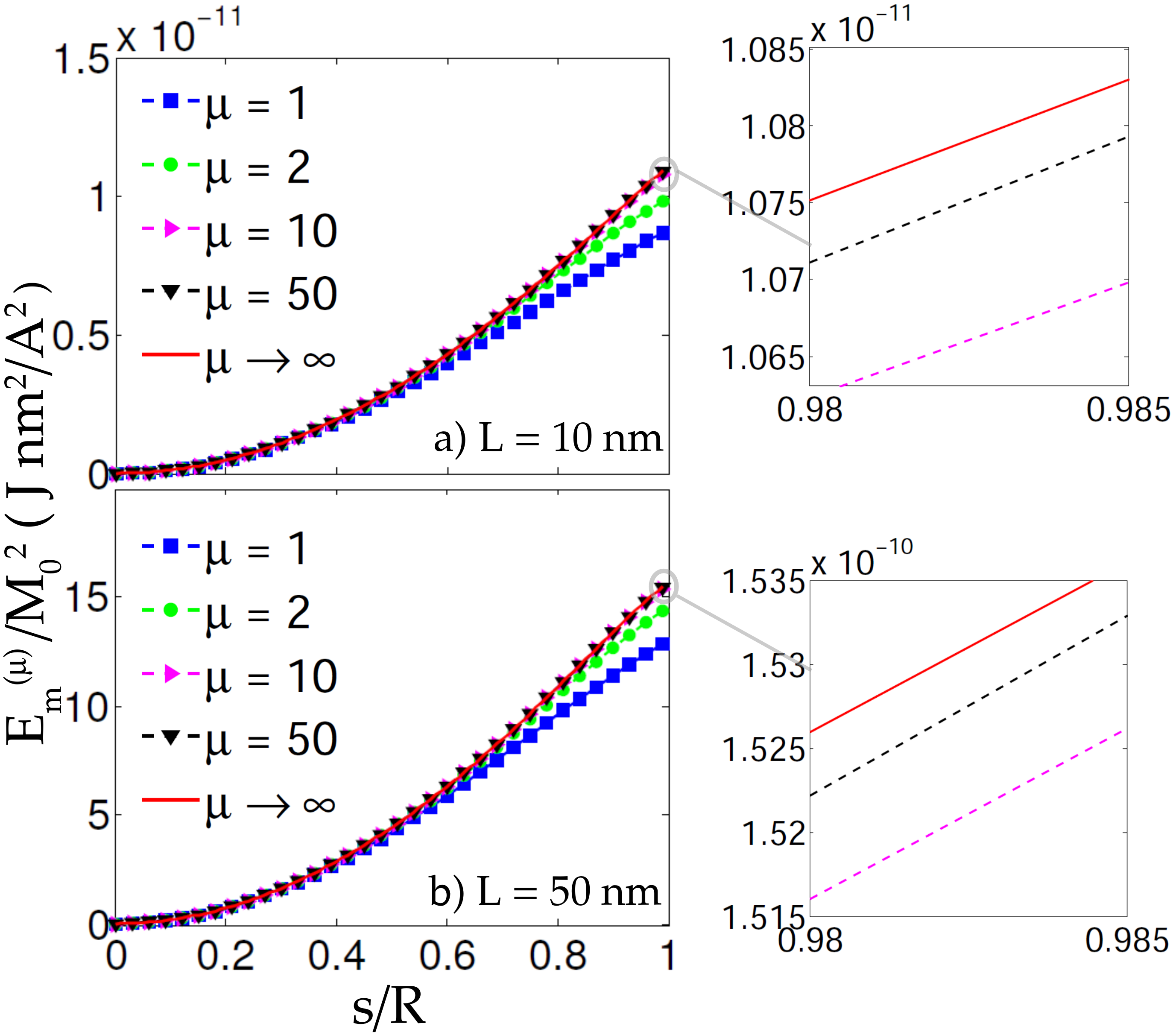}
\caption{(Color online) $E_m^{(\mu)}/M_0^2$ as a function of $s/R$ for a vortex configuration or a dot of $R = 100$ nm in radius and a) $L = 10$ nm and b) $L =50$ nm in thickness. The insets show a zoom for $0.98 \leq s/R \leq 0.985$.}
\label{covergence_Edm}
\end{figure}

\section{Results and discussion}

\begin{figure}[ht]
\centering
\includegraphics[width=1.0\linewidth]{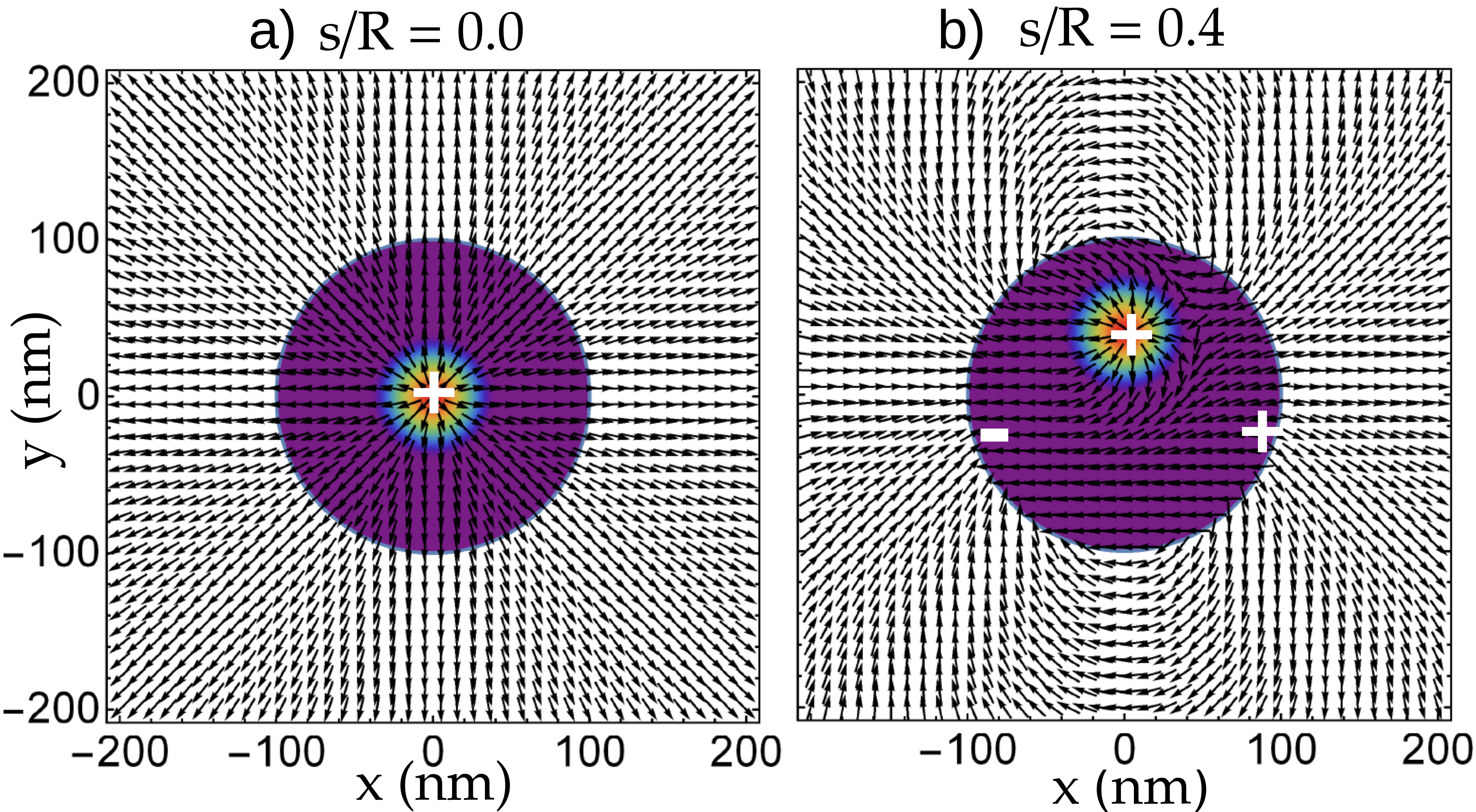}
\caption{(Color online) Magnetostatic field in the $xy$ plane, $\vec{h}_d^{(xy)}$, in a cylinder of $R=100$ nm radius and $L=70$ nm thickness, with a chirality $\mathcal{C} = 1$ and a core radius of $R_c = 40$ nm. (a) shows a vortex without displacement, that is, centered on the cylinder, while (b) shows a displaced vortex core. The $+$ and $-$ signs are guide to the eyes to understand the field lines behavior using the dumbbell picture \cite{CMS07}.}
\label{mag_field_xyplane}
\end{figure}

To better understand the effect of the displacement of the vortex core in the cylinder, we have plotted in Fig. \ref{mag_field_xyplane} the magnetostatic field $\vec{H}_\text{d}= - \vec{\nabla} U$, where $U = U_c + U_m(s)$. In particular, Fig. 3 shows the magnetostatic field in the $xy$ plane (top view of the cylinder), normalized to this plane, that is, $\vec{h}_d^{(xy)} = [(\vec{H}_d \cdot \hat{x} )\,\hat{x} + (\vec{H}_d \cdot \hat{y} )\, \hat{y}]/\sqrt{(\vec{H}_d \cdot \hat{x} )^2+(\vec{H}_d \cdot \hat{y} )^2}$, in a cylinder of $R=100$ nm radius and $L=70$ nm thickness, with a chirality $\mathcal{C} = 1$ and a core radius of $R_c = 40$ nm. Fig. 3a shows a vortex without displacement, that is, centered on the cylinder, while Fig. 3b shows a displaced vortex core. When the core is located in the center of the cylinder, there are no surface charges in the cylinder mantle (see Fig. 3a). However, when the core moves, due to the application of an external magnetic field in the $x$ direction, negative and positive surface charges appear accumulating to the left and right of the cylinder, respectively, because the region with parallel magnetic moments to the $x$ direction increases its size. This arrangement of opposite charges on the edges of the cylinder produces the magnetic field lines shown in Fig. 3b.

As we saw previously, the displacement of the vortex core produces surface charges in the mantle of the cylinder, charges that are associated with the appearance of a surface anisotropy, $E_s = K_s \int_{\mathcal{S}_m} d\mathcal{S} \, (\hat{n} \cdot \hat{m})^2/2$, given by,
\begin{equation}
\label{Es}
E_s = \frac{K_s L R}{2} \frac{s^2}{R^2} \int_0^{2\pi} d\phi \, \frac{\cos^ 2{\phi}}{1 + s^2/R^2 - 2 \sin{\phi} \, s/R  } \, ,
\end{equation}
where $K_s$ corresponds to the surface anisotropy constant. By equating $E_s$ with the expansion in $\mu$ of $E_m^{(\mu)}$ \cite{GNO+01} at second order in $s/R$, we have obtained an expression for $K_s$ given by
\begin{equation}
\label{Ks}
K_s = \mu_0 M_0^2 R \,  N_y(L/R) \, ,
\end{equation}
with $N_y(\gamma) = \int_0^\infty \frac{dt}{t} \left( 1-\frac{1-e^{-\gamma t}}{\gamma t} \right) J_1^2(t)$ being the cylinder in-plane demagnetizing factor \cite{Aharoni90} in the direction of the deviation of the core. In this way, we have not only reached the conclusion that the displacement of the vortex core induces a surface anisotropy in the cylinder, but also, we have obtained a simplified expression, $E_s$, for the dipolar energy of the mantle, $E_m$. It is worthwhile to mention that the results obtained in Eqs.\eqref{Um},\eqref{Edm},\eqref{Es} and \eqref{Ks} are valid for a general vortex Ritz model $\vec{m} = m_z(r) \hat{z} + m_\phi (r) \hat{\phi}$, with $m_\phi = 1 $ for $r \geq R_c$ and $m_\phi = \sqrt{1-m_z(r)^2}$ for $r < R_c$, independently of the particular shape of $m_z(r)$.

\begin{figure}[ht]
\centering
\includegraphics[width=1.0\linewidth]{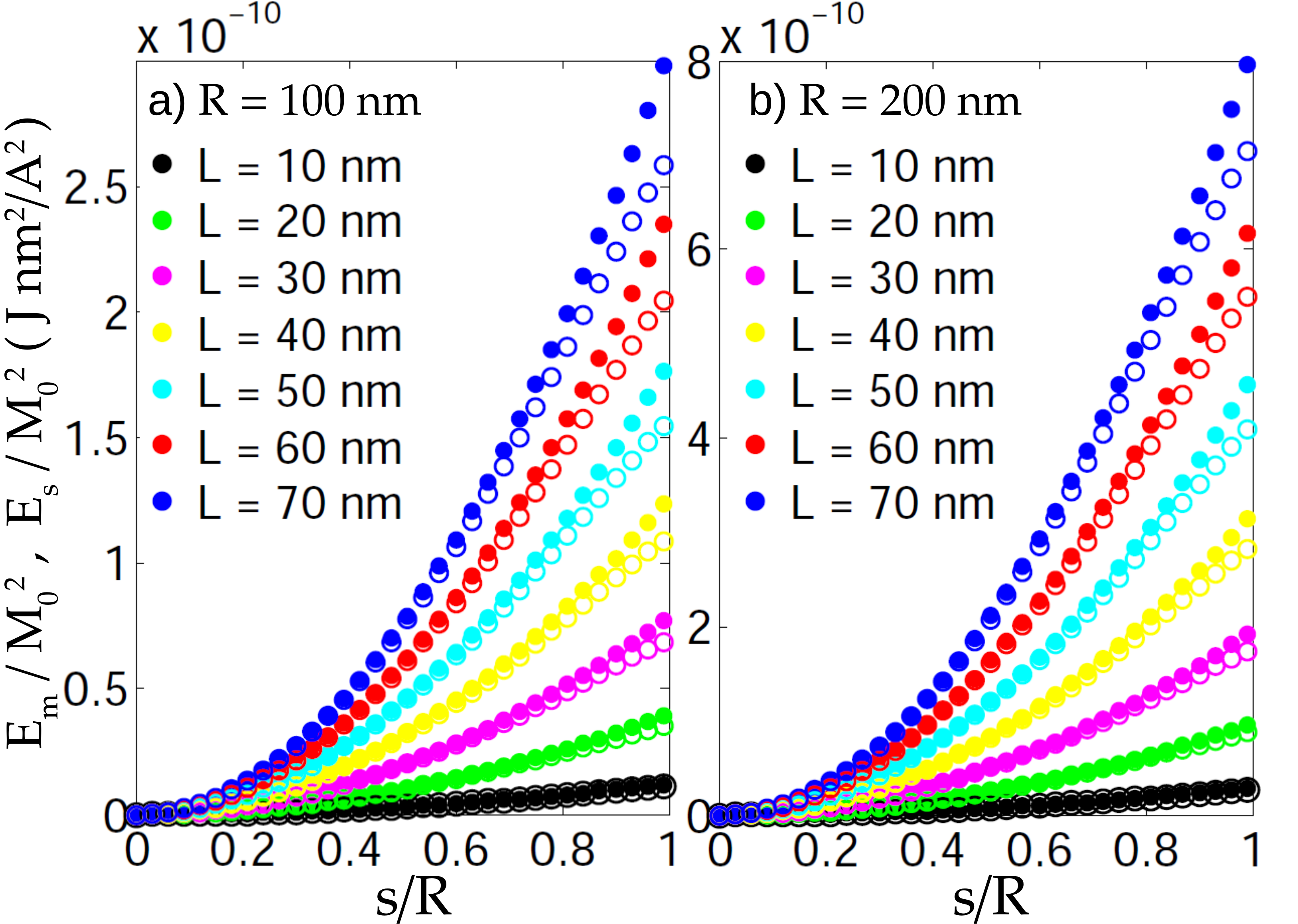}
\caption{(Color online) Comparison of the behavior of $E_s$ given by the Eq. \eqref{Es} with $E_m$ given by the Eq. \eqref{Edm} as a function of $s/R$ for cylinders with different thicknesses $L$ and radii of (a) $R = 100$ nm and (b) $R = 200$ nm. Hollow and solid circles correspond to the $E_m$ and $E_s$ curves, respectively.}
\label{Es_fit_Edm}
\end{figure}

In order to compare $E_s$ with $E_m$, we have plotted in Fig. \ref{Es_fit_Edm} both energy terms  (over $M_0^2$) as a function of $s/R$ for a cylinder of radius (a) $R = 100$ nm and (b) $R = 200$ nm, for different thicknesses $L$. $E_s$ was calculated using Eq. \eqref{Es} with $K_s$ given by the Eq. \eqref{Ks}, which is shown as solid circles, whereas $E_m$ was obtained from Eq. \eqref{Edm} and is represented as hollow circles. The different $L$-curves are distinguished by different colors.  
From Fig. 4 it can be seen that $E_s$ has the same behavior as $E_m$, and that both energy terms are in the same order of magnitude for the entire range of parameters considered, both for $R = 100$ nm and $R = 200$ nm, regardless of the value of the thickness of the cylinder and its magnetic parameters.

The goodness of fit of $E_s$ to the dipolar energy term $E_m$ is estimated for the numerical data using the coefficient of determination $\mathcal{R}^2$,
\begin{equation}
\label{Rsq_coeff}
\mathcal{R}^2 = 1 - \frac{\sum_i \left[ E_s(s_i)- E_m(s_i)\right]^2}{\sum_i \left[ E_m(s_i)- \left< E_m \right> \right]^2} \, ,
\end{equation}
where $\left< E_m \right>$ is the average value of $E_m(s_i)$. We have found a high accuracy fit when $s/R \leq 0.5$ for both radii, independently of the thickness L. The detailed values of the $\mathcal{R}^2$-coefficient are given in Table I. From these data it can be mentioned that, when $s/R \leq 0.5$, the fit explains at least $99.99\%$ of the total variation in the data about the average of $E_m(s_i)$, and the accuracy fit of $E_s$ increases as the dot thickness decreases. Moreover, since the $\mathcal{R}^2$ could be misleading for non linear regression, we have also computed the standard error deviation of the data founding typical errors between 1 per thousand and 1 per 10 thousand, in the scale of Fig. \ref{Es_fit_Edm}.
\begin{table}[ht]
\centering
\label{Rsq_values}
\begin{tabular}{|l|cc|}

\multicolumn{2}{l}{}   \\  \hline

\hspace{0.7cm}$\mathcal{R}^2$ &  $R = 100$ nm & $R = 200$ nm \\ \hline

$L = 10$ nm   & 0.999993     & 0.999997 \\ \hline
$L = 20$ nm  & 0.999988    & 0.999993   \\ \hline
$L = 30$ nm  & 0.999980    & 0.999992   \\ \hline
$L = 40$ nm  & 0.999974    & 0.999988   \\ \hline
$L = 50$ nm  & 0.999968    & 0.999983   \\ \hline
$L = 60$ nm  & 0.999963    & 0.999980   \\ \hline
$L = 70$ nm  & 0.999957    & 0.999976   \\ \hline
\end{tabular}
\caption{Coefficient of determination $\mathcal{R}^2$ of the fit for $s/R \leq 0.5$.}
\end{table}

\begin{figure}[ht]
\centering
\includegraphics[width=0.8\linewidth]{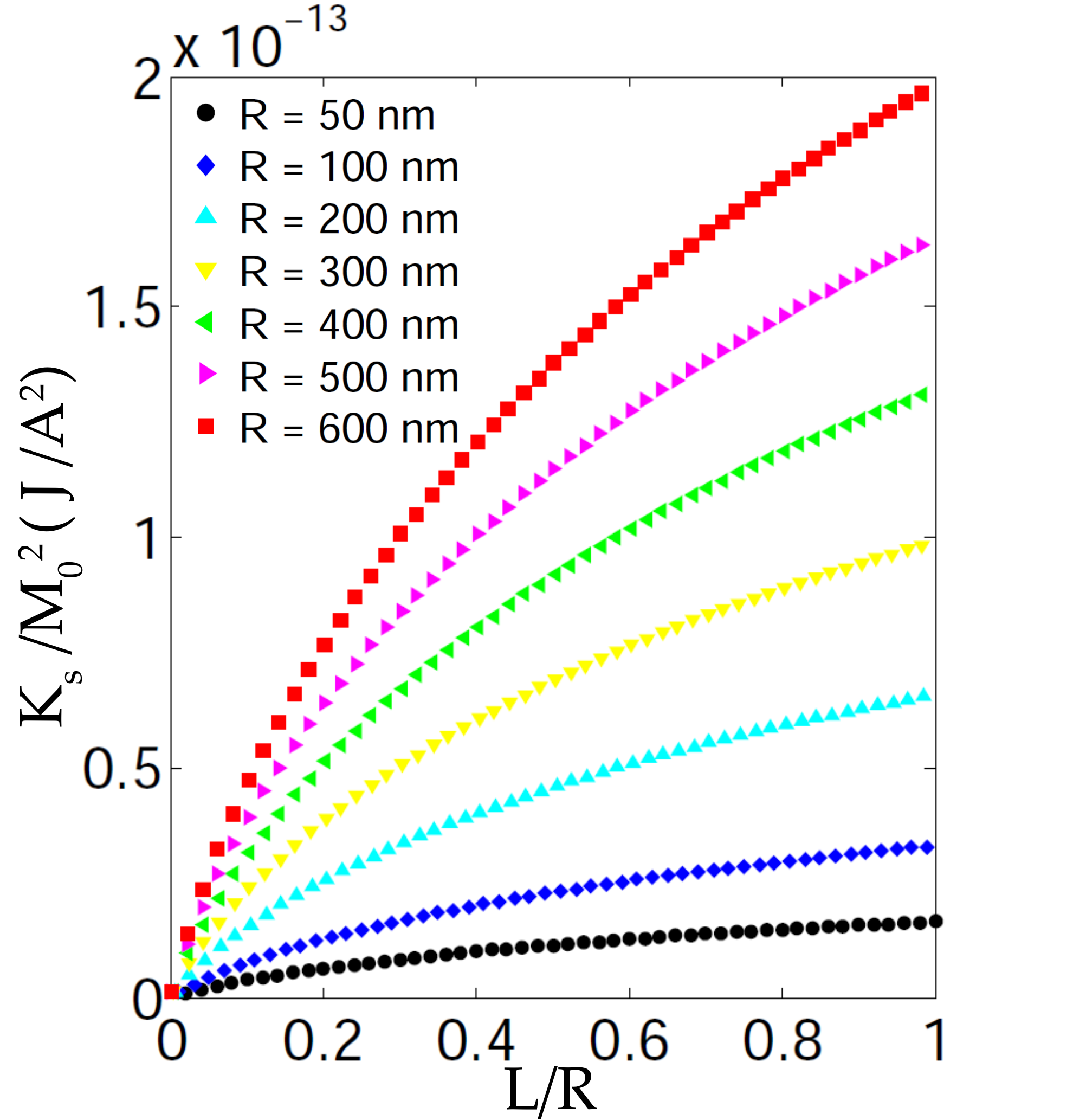}
\caption{(Color online) surface anisotropy constant $K_s$ (over $M_0^2$) as a function of the cylinder aspect ratio, $L/R$, for different radii.}
\label{Ks_Fig}
\end{figure}

In Fig. \ref{Ks_Fig} we show the surface anisotropy constant $K_s$ (over $M_0^2$) as a function of the cylinder aspect ratio, $L/R$, for different radii. This constant has been obtained from Eq. 4. From this figure we can obtain the value of the surface anisotropy constant $K_s$ for cylinders with different magnetic and geometric parameters. It is important to note that $K_s$ increases as both the radius and the aspect ratio of the cylinder increase.

Figure 6 shows the displacement of the core, $s/R$, which minimizes the total energy, $E(s)$, which includes not only the dipolar energy, but also the exchange and Zeeman energies obtained from Ref. [40]. In this case we have considered cobalt cylinders with a saturation magnetization of $M_0 =1.4 \times 10^6$ A/m and an exchange stiffness constant of $A= 3 \times 10^{-11}$ J/m, with radii (a) $R = 100$ nm and (b) $R = 200$ nm, for different thicknesses (to calculate the total energy we have used the particular vortex Ritz model of Ref.\cite{MAL+10}, which was also studied in Refs\cite{RVL+16,RVT+18}). The circles were obtained by considering the contribution of the mantle, $E_m$, while the lines were obtained using the approximation of the surface energy, $E_s$. When the vortex is the stable magnetic configuration, the circles are solid, whereas when a uniform state is stable (and therefore the vortex is unstable), the circles are hollow. It can be seen a very good agreement between circles and lines data when the vortex state is stable.

\begin{figure}[ht]
\centering
\includegraphics[width=1.0\linewidth]{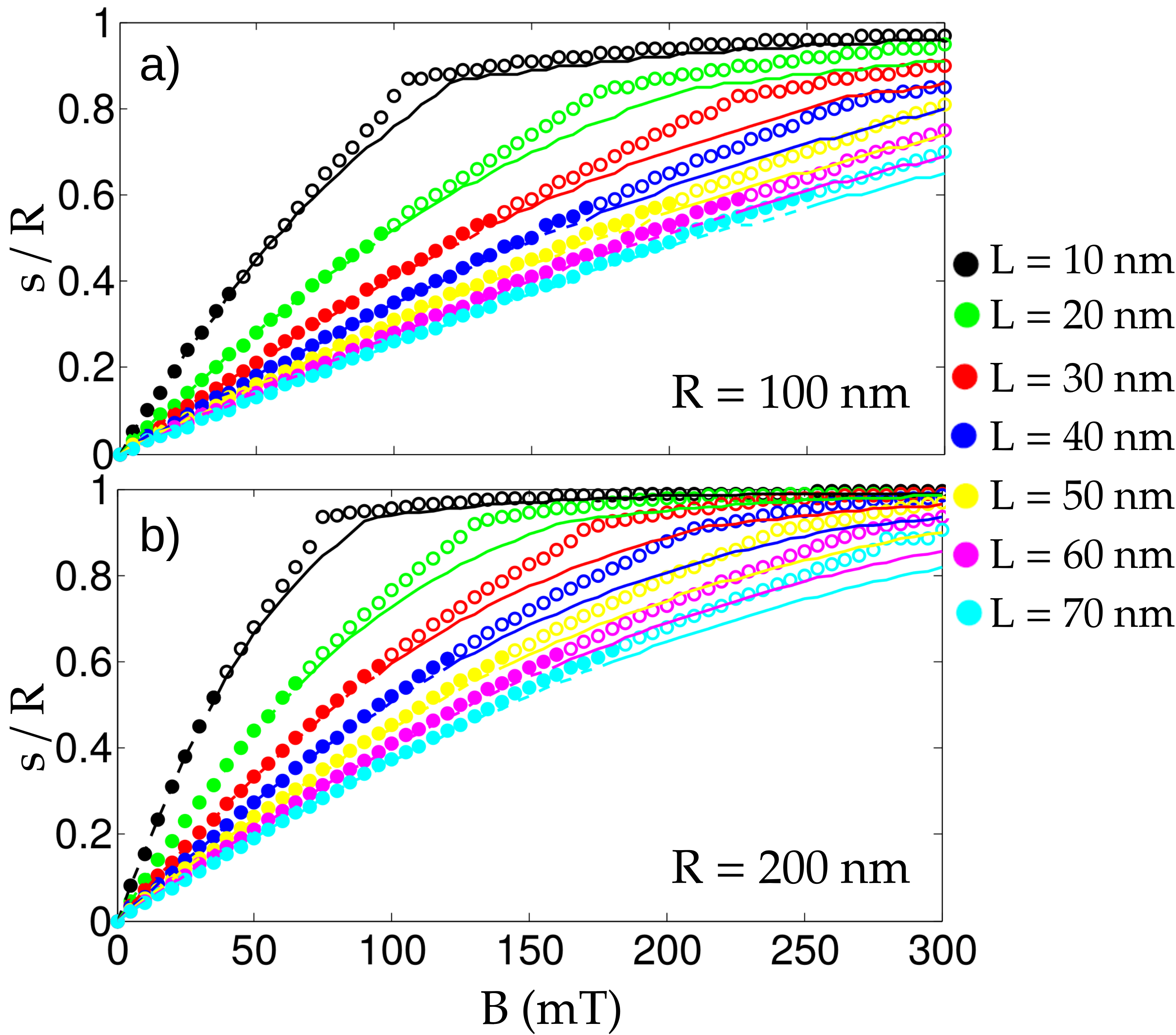}
\caption{(Color online) Displacement of the core, $s/R$, which minimizes the total energy, $E(s)$, as a function of the intensity of the in-plane applied magnetic field, for cobalt cylinders with radii (a) $R = 100$ nm and (b) $R = 200$ nm, for different thicknesses. The circles were obtained by considering the contribution of the mantle, $E_m$, while the lines were obtained using the approximation of the surface energy, $E_s$.}
\label{s_vs_B_Fig}
\end{figure}

\section{Conclusions}
In conclusion, we have shown that the effect of the displacement of the vortex core in the dipolar energy can be modeled simply as a surface anisotropy of the form $E_s = K_s \int_{\mathcal{S}_m} d\mathcal{S} \, (\hat{n} \cdot \hat{m})^2/2$. Interestingly, $E_s$ estimates the behavior of dipolar energy quite well over the entire range of parameters investigated, and in particular, with high accuracy for $s/R \leq 0.5$. Moreover, we have found that the surface constant is proportional to the cylinder in-plane demagnetizing factor in the direction of the core deviation, $N_y(L/R)$, i.e., $K_s = \mu_0 M_0^2 R \,  N_y(L/R)$, where $R$ and $L$ are the radius and the thickness of the cylinder, respectively. This expression can be used to calculate the surface constant of cylinders with different magnetic and geometrical parameters.         

\section*{Acknowledgements}
This work was supported by the Fondecyt Grants 1150952 and 3180470, and Financiamiento Basal para Centros Cientificos y Tecnologicos de Excelencia FB0807.

\end{document}